\documentclass[twocolumn]{aastex631}

\usepackage{amsmath}
\usepackage{amssymb}
\usepackage{natbib}
\usepackage{enumerate}
\usepackage{hyperref}

\usepackage{savesym}
\savesymbol{tablenum}
\usepackage{siunitx}
\newcommand{\imag}{\operatorname{Im}}

\newcommand{\ii}{\mathrm{i}}

\newcommand{\diff}[1]{\operatorname{d}\!{#1}}
\newcommand{\deriv}[3][0]{\frac{\operatorname{d}\if#10\else^{#1}\!\fi\!{#2}}{\operatorname{d}\!{#3}\if#10\else^{#1}\fi}}
\newcommand{\sderiv}[3][0]{\operatorname{d}\if#10\else^{#1}\fi\!{#2}/\operatorname{d}\!{#3}}
\newcommand{\pderiv}[3][0]{\frac{\partial\if#10\else^{#1}\!\fi{#2}}{\partial{#3}}}
\newcommand{\pcderiv}[4][0]{\left(\frac{\partial\if#10\else^{#1}\!\fi{#2}}{\partial{#3}}\right)_{#4}}
\newcommand{\spderiv}[3][0]{\partial\if#10\else^{#1}\fi{#2}/\partial{#3}}

\newcommand{\ver}{\mathbf{e}_{r}}
\newcommand{\vetheta}{\mathbf{e}_{\theta}}
\newcommand{\vephi}{\mathbf{e}_{\phi}}

\newcommand{\vpos}{\mathbf{r}}
\newcommand{\vxi}{\boldsymbol{\xi}}

\newcommand{\txir}[1][0]{\tilde{\xi}_{{\rm r}\if#10\else;#1\fi}}
\newcommand{\txih}[1][0]{\tilde{\xi}_{{\rm h}\if#10\else;#1\fi}}

\newcommand{\tP}[1][0]{\tilde{P}_{\if#10\else#1\fi}}
\newcommand{\trho}[1][0]{\tilde{\rho}_{\if#10\else#1\fi}}
\newcommand{\tPhi}[1][0]{\tilde{\Phi}_{\if#10\else#1\fi}}

\newcommand{\proj}[1][0]{\mathcal{P}_{\if#10\else#1\fi}}
\newcommand{\weight}[1][0]{\mathcal{W}_{\if#10\else#1\fi}}
\newcommand{\surf}[1][0]{\mathcal{S}_{\if#10\else#1\fi}}

\newcommand{\npoly}{n_{\rm poly}}

\newcommand{\polyz}{\xi}

\newcommand{\polyt}{\vartheta}

\newcommand{\brunt}{\mathcal{N}}

\newcommand{\xb}{x_{\rm b}}

\newcommand{\sigmae}{\sigma^{\rm e}}
\newcommand{\sigmai}{\sigma^{\rm i}}

\newcommand{\err}{\mathcal{E}}
\newcommand{\errfreq}[1][0]{\err^{\sigma}_{\if#10\else#1\fi}}
\newcommand{\errorthog}[1][0]{\err^{\proj}_{\if#10\else#1\fi}}
\newcommand{\errfirst}[1][0]{\err^{I}_{\if#10\else#1\fi}}

\newcommand{\defect}[1][0]{\mathcal{D}^{\if#10\else#1\fi}}

\newcommand{\order}[1]{O(#1)}

\newcommand{\adisc}{a_{\rm d}}
\newcommand{\bdisc}{b_{\rm d}}

\newcommand{\aquad}{a_{\rm q}}
\newcommand{\bquad}{b_{\rm q}}

\newcommand{\ainterp}{a_{\rm i}}
\newcommand{\binterp}{b_{\rm i}}

\newcommand{\acont}{a_{\rm c}}
\newcommand{\bcont}{b_{\rm c}}
  
\newcommand{\arandom}{a_{\rm r}}

\newcommand{\gyre}{GYRE}
\newcommand{\mad}{MAD}
\newcommand{\mesa}{MESA}
\newcommand{\nosc}{NOSC}
\newcommand{\adipls}{ADIPLS}
\newcommand{\cafein}{CAFEIN}
\newcommand{\osc}{OSC}
\newcommand{\pulse}{PULSE}
\newcommand{\graco}{GraCo}
\newcommand{\filou}{filou}
\newcommand{\lnawenr}{LNAWENR}
\newcommand{\odepack}{ODEPACK}
\newcommand{\lppul}{LP-PUL}

\newcommand{\Msun}{M_{\odot}}
\newcommand{\Lsun}{L_{\odot}}

\newcommand{\bvtext}{Brunt-V\"ais\"al\"a}

\graphicspath{{./}{figures/}}

\begin{document}

\title{Tools for Characterizing the Numerical Error of Stellar
  Oscillation Codes}

\author[0000-0002-2522-8605]{R. H. D. Townsend}
\affiliation{Department of Astronomy, University of Wisconsin-Madison, 475 N Charter St, Madison, WI 53706, USA}
\affiliation{Center for Computational Astrophysics, Flatiron Institute, New York, NY 10010, USA}
\author[0009-0003-6919-5221]{R. V. Kuenzi}
\affiliation{Department of Astronomy, University of Wisconsin-Madison, 475 N Charter St, Madison, WI 53706, USA}
\author[0000-0001-5137-0966]{J. Christensen-Dalsgaard}
\affiliation{Department of Physics and Astronomy, Aarhus University, Ny Munkegade 120, 8000 Aarhus C, Denmark}

\begin{abstract}
Stellar oscillation codes are software instruments that evaluate the
normal-mode frequencies of an input stellar model. While inter-code
comparisons are often used to confirm the correctness of calculations,
they are not suitable for characterizing the numerical error of an
individual code. To address this issue, we introduce a set of tools
--- `error measures' --- that facilitate this characterization. We
explore the behavior of these error measures as calculation
parameters, such as the number of radial grid points used to
discretize the oscillation equations, are varied; and we summarize
this behavior via an idealized error model. While our analysis focuses
on the \gyre\ code, it remains broadly applicable to other oscillation
codes.
\end{abstract}


\correspondingauthor{Rich Townsend}
\email{townsend@astro.wisc.edu}

\section{Introduction} \label{s:intro}

Recent space photometry missions, including \emph{MOST}
\citep{Walker:2003}, \emph{CoRoT} \citep{Baglin:2006}, \emph{Kepler}
\citep{Borucki:2010}, and \emph{TESS} \citep{Ricker:2014}, have
provided time-series measurements of oscillations in myriad stars
spanning the Hertzsprung-Russell diagram. The technique of
asteroseismology \citep[e.g.,][]{Aerts:2010} can leverage these
data into quantitative constraints on the global
properties (mass, radius, age, metallicity, etc.) and detailed
interior structures of the stars observed.

A key ingredient in asteroseismic modeling is a stellar oscillation
code, a software instrument that evaluates the normal-mode frequencies
of an input stellar model. Many such codes are described in the
literature; examples include \mad\ \citep{Dupret:2001},
\lppul\ \citep{Corsico:2002}, \pulse\ \citep{Brassard:2008},
\adipls\ \citep{Christensen-Dalsgaard:2008},
\graco\ \citep{Moya:2008a}, \nosc\ \citep{Provost:2008},
\osc\ \citep{Scuflaire:2008}, \filou\ \citep{Suarez:2008},
\lnawenr\ \citep{Suran:2008}, \cafein\ \citep{Valsecchi:2013}, and
\gyre\ \citep{Townsend:2013,Townsend:2018,Goldstein:2020,Sun:2023}. Given
the community's increasing reliance on these complex software
packages, it is important to verify that they yield results that are
not only precise but also accurate.

Historically, a number of papers have examined the problem of code
verification. In considering the oscillations of polytropic models,
\citet{Mullan:1988} employ a convergence criterion that the change in
mode frequencies, upon increasing the number of radial grid points used to
discretize the oscillation equations, drops below a prescribed
threshold. Such an approach can ensure that frequencies are precise,
but does not by itself also guarantee accuracy. In a subsequent paper,
\citet{Christensen-Dalsgaard:1994} expand on this work by undertaking
a comprehensive comparison between the polytrope-model frequencies
evaluated by a pair of independent oscillation codes. More recently,
\citet{Moya:2008b} compare many of the codes cited above, as applied
to realistic stellar models.

Inter-code comparisons provide confidence that oscillation codes are
yielding reliable results. However, their utility is limited when it
comes to characterizing the numerical error of an individual code ---
for instance, diagnosing the various mechanisms that contribute toward
this error, and exploring whether these mechanisms behave as expected
from theoretical considerations. Such characterizations can
potentially highlight hidden bugs, and are key to deciding if and how
the code can further be improved.

Within this context, the present paper introduces a set of tools for
characterizing the numerical error of oscillation codes. We introduce
these tools --- `error measures' --- in Section~\ref{s:error}, and
demonstrate their application to the \gyre\ code in
Section~\ref{s:calcs}. We summarize our findings in
Section~\ref{s:discuss}, and discuss some of the practicalities of
using the tools in the context of asteroseismic studies.


\section{Error measures} \label{s:error}

The numerical error of an oscillation code can in principle be
quantified by comparing the code's outputs (eigenfrequencies,
eigenfunctions, etc.) against exact solutions. In practice, however,
the latter are known only for the simple and quite artificial case of
the homogeneous compressible sphere \citep[see][]{Pekeris:1938}. To
proceed, therefore, we heed these words from
\citet{Christensen-Dalsgaard:1991}: ``The only practical way of
estimating the numerical error in the computed frequencies is to
compare results obtained with different numbers of mesh [grid]
points''.

The general process is straightforward: run a code multiple times
while varying the number of radial grid points $N$, and explore how
its outputs change with $N$. As a straightforward example, one might
follow \citet{Mullan:1988} and focus on how the frequency $\sigma$ of
a given mode depends on $N$. However, as we already mention above,
this approach ensures precision but does not guarantee accuracy. The
problem lies with the choice of $\sigma$ as the output to monitor; by
itself, it furnishes no indication that it lies close to the
correct value. To address this, we introduce the concept of an error
measure $\err$, fulfilling the following criteria:
\begin{enumerate}[(i)]
\item $\err$ must be calculable from the outputs of the oscillation
  code; \label{i:prop-eval}
\item $\err$ must be insensitive to non-physical quantities such as
  the normalization of eigenfunctions; \label{i:prop-norm}
\item $\err$ must approach zero as numerical solutions approach exact
  solutions of the oscillation equations. \label{i:prop-zero}
\end{enumerate}
The last criterion is key to ensuring that precision and accuracy go
hand-in-hand. In the subsections that follow, we introduce three
specific definitions of the error measure. To keep the presentation
compact, we defer much of the detailed mathematical formalism to
Appendix~\ref{a:formal}.

\begin{figure*}[ht]
  \includegraphics{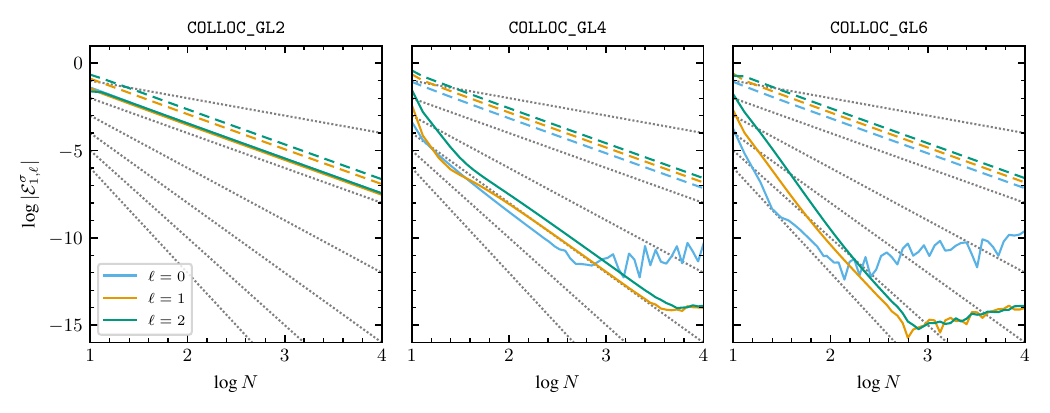}
  \caption{The frequency error measure $\errfreq[n,\ell]$ plotted as a
    function of the number of grid points $N$, for $n=1$, $\ell=0,1,2$
    modes of the $\npoly=1$ polytrope. The panels show differing
    choices of the discretization scheme, indicated at
    top. Solid (dashed) lines show data evaluated using the
    Newton-Coates (trapezoidal) quadrature schemes.  To help asses the
    rate of convergence, the dotted lines indicate $|\errfreq[n,\ell]| =
    N^{-j}$ for $j=1,\ldots,6$.} \label{f:anapoly-1-err-freq}
\end{figure*}

\begin{figure*}[ht]
  \includegraphics{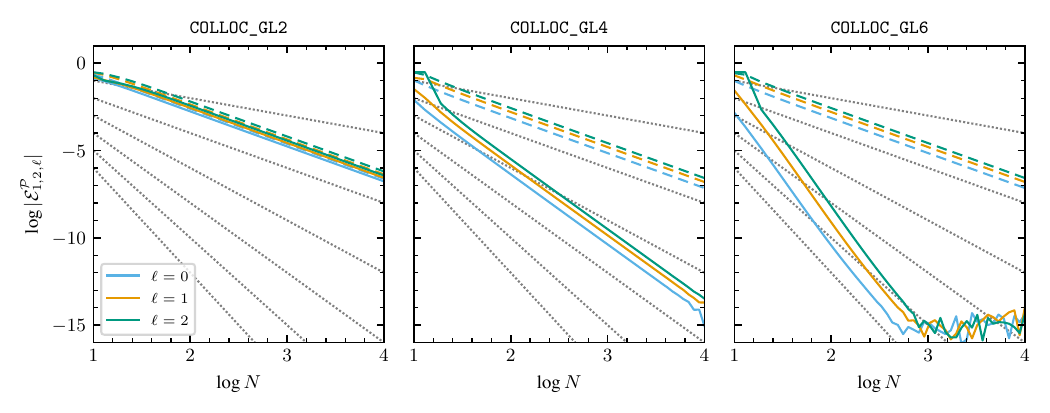}
  \caption{As in Fig.~\ref{f:anapoly-1-err-freq}, except that the
    orthogonality error measure $\errorthog[n,n',\ell]$ is now plotted
    for $(n,n')=(1,2)$ modes.} \label{f:anapoly-1-err-orthog}
\end{figure*}

\begin{figure*}[ht]
  \includegraphics{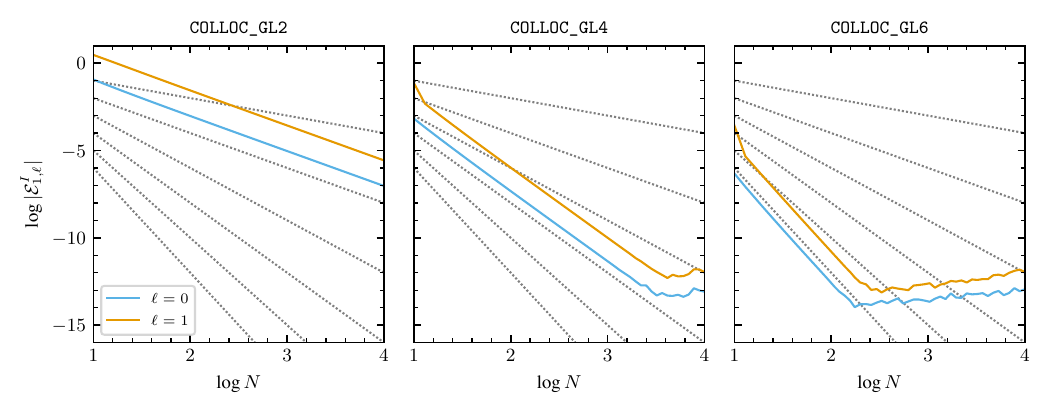}
  \caption{As in Fig.~\ref{f:anapoly-1-err-freq}, except that the
    first-integral error measure $\errfirst[n,\ell]$ is now plotted
    for $n=1$ modes. Note that this error measure is defined only for
    $\ell=0,1$, and thus no $\ell=2$ curve is
    shown.} \label{f:anapoly-1-err-first}
\end{figure*}

\begin{figure*}[ht]
  \includegraphics{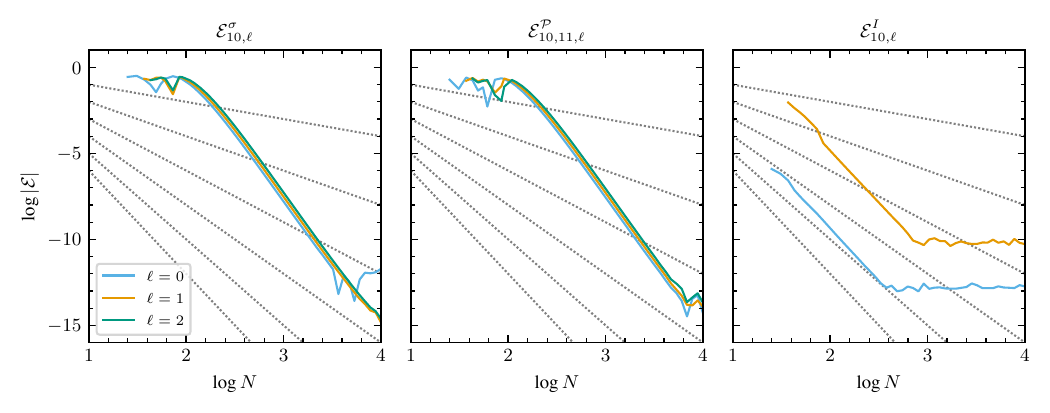}
  \caption{The $\errfreq[n,\ell]$, $\errorthog[n,n',\ell]$ and
    $\errfirst[n,\ell]$ error measures plotted as a function of the
    number of grid points $N$, for $(n,n')=(10,11)$, $\ell=0,1,2$
    modes of the $\npoly=1$ polytrope. All calculations use the
    \texttt{COLLOC\_GL6} discretization scheme. The dotted lines are
    the same as in previous figures. No data are shown at small $N$
    because \gyre\ is unable to find the requisite modes on such coarse
    grids.} \label{f:anapoly-1-high-order-err}
\end{figure*}

\subsection{Frequency error measure} \label{s:error-freq}

As discussed in Appendix~\ref{a:freq}, the frequency of a mode can be
evaluated from various integrals over its eigenfunctions. For
numerical solutions, this `integral' frequency will not agree exactly
with the eigenfrequency output by the oscillation code, and we
introduce the frequency error measure
\begin{equation} \label{e:error-freq}
\errfreq[n,\ell] \equiv \frac{\sigmae_{n,\ell} - \sigmai_{n,\ell}}{\sigmae_{n,\ell}}
\end{equation}
to quantify their relative difference. Here, $\sigmae_{n,\ell}$ is the
eigenfrequency output by the code, and $\sigmai_{n,\ell}$ is the
integral frequency evaluated using equation~(\ref{e:freq});
throughout, the subscripts $n$ and $\ell$ denote the radial order and
harmonic degree, respectively, of the mode.

\subsection{Orthogonality error measure} \label{s:error-orthog}

As discussed in Appendix~\ref{a:orthog}, the displacement
eigenfunctions of a pair of oscillation modes with the same harmonic
degree but different radial orders ($n,n'$) are orthogonal if the
surface term~(\ref{e:surf-term}) satisfies $\surf[n,n',\ell] =
\surf[n',n,\ell]^{\ast}$. Numerical solutions will not satisfy this
property exactly, and we introduce the orthogonality
error measure
\begin{equation} \label{e:error-orthog}
  \errorthog[n,n',\ell] \equiv \frac{\proj[n,n',\ell]}{\left( \proj[n,n,\ell] \, \proj[n',n',\ell] \right)^{1/2}},
\end{equation}
to quantify the discrepancy; here, the inner product
  $\proj[n,n',\ell]$ is defined in equation~(\ref{e:proj}). In
accordance with criterion~(\ref{i:prop-norm}), above, the denominator
ensures that this measure remains independent of the arbitrary
normalization of eigenfunctions.

\subsection{First-Integral Error Measure} \label{s:error-first}

As shown by \citet{Takata:2006a}, the oscillation equations for radial
($\ell=0$) and dipole ($\ell=1$) modes admit first integrals that
vanish everywhere for solutions that remain regular at the
origin. Once again, numerical solutions to the equations will not
satisfy this property exactly, and we introduce the first-integral
error measure
\begin{equation} \label{e:error-first}
  \errfirst[n,\ell] \equiv \frac{\max(I_{n,\ell}) - \min(I_{n,\ell})}{\proj[n,n,\ell]^{1/2}}
\end{equation}
to quantify the discrepancy. Here, $I_{n,\ell}$ ($\ell = 0,1$) are the
first integrals defined in equations~(42) and~(43) of
\citet{Takata:2006a} for radial and dipole cases, respectively; and
the $\min()$ and $\max()$ operators denote the minimum and maximum
value attained over all points of the radial grid. As with
equation~(\ref{e:error-orthog}), the denominator ensures
that this measure is independent of eigenfunction normalization.


\section{Calculations} \label{s:calcs}

For a variety of stellar models, we explore how the error measures
introduced above behave as a function of the number of radial grid
points $N$. We use release 8.0 of the \gyre\ code, which
  includes support for the evaluated models discussed in
  Section~\ref{s:eval-models}. At the stellar surface $r=R$ we impose
the zero-pressure mechanical boundary condition
\begin{equation} \label{e:zero-press-bound}
  \delta \tP[n,\ell] \equiv \tP[n,\ell]' - g \, \rho \, \txir[n,\ell] = 0,
\end{equation}
together with the potential boundary condition
\begin{equation} \label{e:pot-bound}
  \deriv{\tPhi[n,\ell]'}{r} + \frac{\ell+1}{r} \tPhi[n,\ell]' = - 4 \pi \, G \, \rho \, \txir[n,\ell]
\end{equation}
\citep[see, e.g.,][]{Cox:1980}. Together, these boundary conditions
mean that $\surf[n,n',\ell] = \surf[n',n,\ell]^{\ast}$ (see
equation~\ref{e:orthog-with-pot-bound}), and thus eigenfunctions
should be orthogonal. We briefly discuss the impact of other
  boundary-condition choices in Section~\ref{s:discuss}.

With the exception of a few special cases that we highlight when
necessary, the grids employed by \gyre\ consist of $N = 6K + 1$ points
uniformly distributed in fractional radius $0 \leq x
\equiv r/R \leq 1$, where $K$ is an integer. This choice allows us to
evaluate integrals (for instance, those appearing in
equations~\ref{e:freq} and~\ref{e:orthog}) using composite 7-point
Newton-Coates quadrature \citep{Press:1992} across $K$
sub-intervals. The motivation here is to ensure that the quadrature
truncation error remains small in comparison to the other sources of
numerical error arising in the calculations.

\subsection{Evaluated models} \label{s:eval-models}

We first consider stellar models where the internal structure can be
evaluated from closed-form expressions. The most simple are the
polytropes with indices $\npoly=0$, 1 and 5; in these cases,
\gyre\ calculates the structure of the model on-the-fly using the
analytic solutions to the Lane-Emden equation \citep[see,
  e.g.,][]{Horedt:2004}. Setting aside the $\npoly=0$ case (which
corresponds to the homogeneous compressible sphere, and as discussed
above is special), in this section we focus on the $\npoly=1$ and
$\npoly=5$ polytropes, together with a composite polytrope formed from
a combination of these.

\subsubsection{$\npoly=1$ polytrope} \label{s:poly-1-model}

We begin by exploring the behavior of the error measures in the case
of an $\npoly=1$ polytrope. For this, and for all other polytrope
models considered, we adopt a canonical first adiabatic index
$\Gamma_{1} = 5/3$. Figure~\ref{f:anapoly-1-err-freq} plots the
frequency error measure $\errfreq[n,\ell]$ as a function of $N$, for
$n=1$, $\ell=0,1,2$ modes\footnote{Here and throughout, mode radial
orders $n$ follow the Eckart-Osaki-Scuflaire classification scheme
\citep[e.g.,][]{Unno:1989}, as extended for dipole modes by
\citet{Takata:2006b}.}. The three panels correspond to different
choices of the numerical scheme used to discretize the oscillation
equations onto the radial grid. \gyre\ implements a multiple-shooting
approach where the relationship between dependent variables at
adjacent grid points is treated as an initial-value problem (IVP). The
\texttt{COLLOC\_GL{i}} schemes ($i=2,4,6$) referenced in the figure solve
this IVP using Gauss-Legendre collocation
\citep[e.g.,][]{Ascher:1995}, an implicit Runge-Kutta method with
$i/2$ intermediate stages.

The frequency error measures plotted in each panel display broadly
similar behavior. Initially, they decay rapidly with increasing $N$,
asymptoting toward a scaling $\errfreq[n,\ell] \sim N^{-i}$. This
continues until $\errfreq[n,\ell]$ drops to $\approx 10^{-11}$
($\approx 10^{-14}$) for the radial (respectively, non-radial) modes;
beyond these limits, the error measures grow approximately linearly with
$N$, with significant variability from one value to the next.

The asymptotic-decay regime arises when the truncation error of the
discretization scheme makes the dominant contribution toward
$\errfreq[n,\ell]$. Theoretically, the local truncation error of the
\texttt{COLLOC\_GL}$i$ scheme is $\order{h^{i+1}}$, where $h =
(N-1)^{-1} \approx N^{-1}$ is the spacing between adjacent grid
points. This yields a global error scaling as $\sim N (N-1)^{-i-1}
\approx N^{-i}$, as seen in each panel of the figure.

The switch to the slow-growth regime occurs when the dominant
contribution toward $\errfreq[n,\ell]$ transitions to round-off
error. \gyre\ performs all calculations using IEEE 754 binary64
floating-point arithmetic, which has a 53-bit significand and
therefore suffers a relative rounding error of up to $2^{-53}
\approx \num{e-16}$ during each arithmetic operation. The round-off
errors from a sequence of arithmetic operations accumulate
approximately in proportion to the operation count, which in \gyre's
case scales with $N$ --- hence, the approximately linear growth of the
error measures plotted in the figure, once accumulated round-off error
becomes dominant.

The dashed lines in Fig.~\ref{f:anapoly-1-err-freq} show the
consequence of using trapezoidal quadrature to evaluate the integrals
in equation~(\ref{e:freq}), instead of the Newton-Coates scheme
described previously. While the output from \gyre\ remains unaltered
by this choice, the frequency error measure in the asymptotic-decay
regime now behaves as $\errfreq[n,\ell] \sim N^{-2}$, because the
$\order{h^{3}}$ local truncation error of trapezoidal quadrature
dominates the truncation error of the discretization schemes [for
  comparison, the local truncation error of the Newton-Coates
  quadrature scheme is $\order{h^{9}}$].

Fig.~\ref{f:anapoly-1-err-freq} confirms that \gyre\ performs as
expected, although the radial mode seems to be considerably more
affected by round-off error than the other modes, which warrants
further investigation. Continuing our analysis,
Figs.~\ref{f:anapoly-1-err-orthog} and~\ref{f:anapoly-1-err-first}
reprise Fig.~\ref{f:anapoly-1-err-freq} by plotting
$\errorthog[1,2,\ell]$ and $\errfirst[1,\ell]$, respectively, as a
function of $N$. In each case, the same general behavior as before is
seen: error measures decaying rapidly with the expected asymptotic
scalings, and then slowly growing when round-off error takes over.

So far we focus on $n=1$ modes, whose eigenfunctions vary slowly with
radius. As a final look at the $\npoly=1$ polytrope,
Fig.~\ref{f:anapoly-1-high-order-err} reprises the right-hand
(\texttt{COLLOC\_GL6}) panels of the previous figures, but now
considering modes with radial orders $(n,n')=(10,11)$ (in the
interests of brevity, we no longer show the results from calculations
using the \texttt{COLLOC\_GL2} or \texttt{COLLOC\_GL4} schemes). The
same general behavior as before is seen, but for the frequency and
orthogonality error measures the approach toward the asymptotic $\err
\sim N^{-6}$ scaling is delayed until larger $N$ is reached. This is
a consequence of the more-oscillatory eigenfunctions of the high-order
modes; smaller $h$ is needed to resolve the eigenfunctions and realize
the expected $\order{h^{7}}$ local truncation error of the
discretization scheme.

\subsubsection{$\npoly=5$ Polytrope} \label{s:poly-5-model}

\begin{figure*}[ht]
  \includegraphics{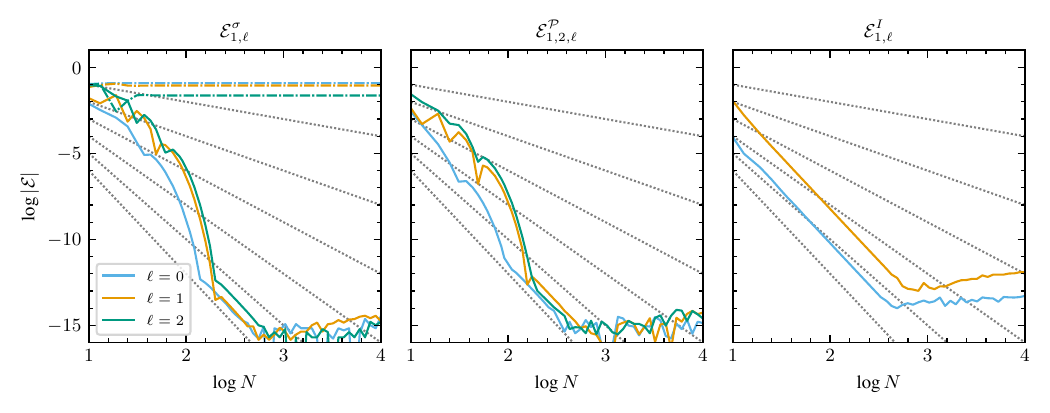}
  \caption{As in Fig.~\ref{f:anapoly-1-high-order-err}, but now for
    $(n,n') = (1,2)$, $\ell=0,1,2$ modes of the $\npoly=5$
    polytrope. In the left-hand panel, the dot-dashed lines show the
    impact of neglecting the surface term $\surf[n,n,\ell]$ term in
    equation~(\ref{e:error-freq}).} \label{f:anapoly-5-err}
\end{figure*}

To explore the impact of a finite surface density, we now consider an
$\npoly=5$ polytrope truncated at the point where the Lane-Emden
dependent variable reaches the value $\polyt=0.2$. This implies a
density $\rho_{\rm s} = 0.2^{5} \,\rho_{\rm c}$ at the stellar surface
(where $\rho_{\rm c}$ is the central density), meaning that the
surface term $\surf[n,n',\ell]$ does not generally vanish ---
although, with the zero-pressure boundary
condition~(\ref{e:zero-press-bound}), it remains the case that
$\surf[n,n',\ell] = \surf[n',n,\ell]^{\ast}$ (see the brief discussion
following equation~\ref{e:orthog-with-pot-bound}).

For this model, Fig.~\ref{f:anapoly-5-err} reprises the
\texttt{COLLOC\_GL6} panels of
Figs.~\ref{f:anapoly-1-err-freq}--\ref{f:anapoly-1-err-first}.  The
error measures are in line with the expectations established by the
$\npoly=1$ case, although in the left-hand and center panels the $\err
\sim N^{-6}$ asymptotic scalings are not reached until $N \gtrsim
100$. In the left-hand panel, the dot-dashed lines reveal the
consequence of neglecting the surface term: once $N \gtrsim 30$,
$\errfreq[n,\ell]$ shows no appreciable change with further increase
in $N$, indicating that the defect introduced by this neglect has
become the dominant contributor toward the frequency error measure.

\subsubsection{Composite $\npoly=5,1$ polytrope} \label{s:poly-5-1-model}

\begin{figure*}[ht]
  \includegraphics{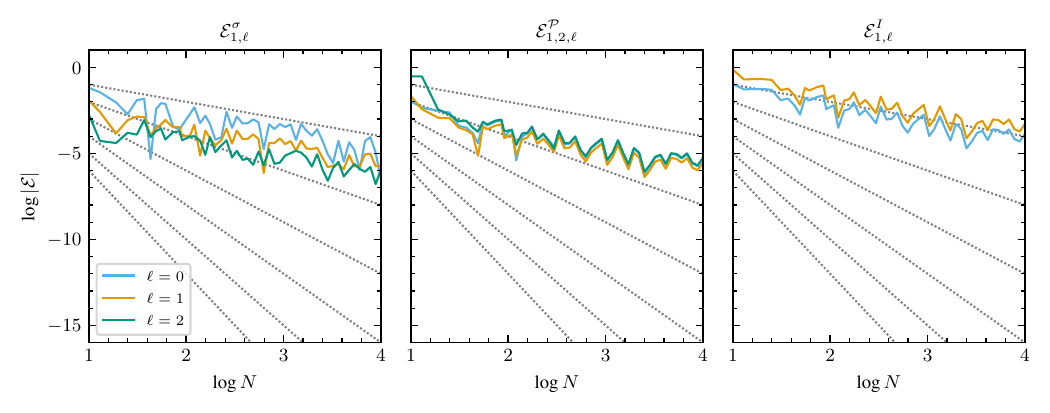}
  \caption{As in Fig.~\ref{f:anapoly-5-err}, but now for
    $(n,n')=(1,2)$, $\ell=0,1,2$ modes of the $\npoly=5,1$ composite
    polytrope.} \label{f:anapoly-5-1-no-resolve-err}
\end{figure*}

\begin{figure*}[ht]
  \includegraphics{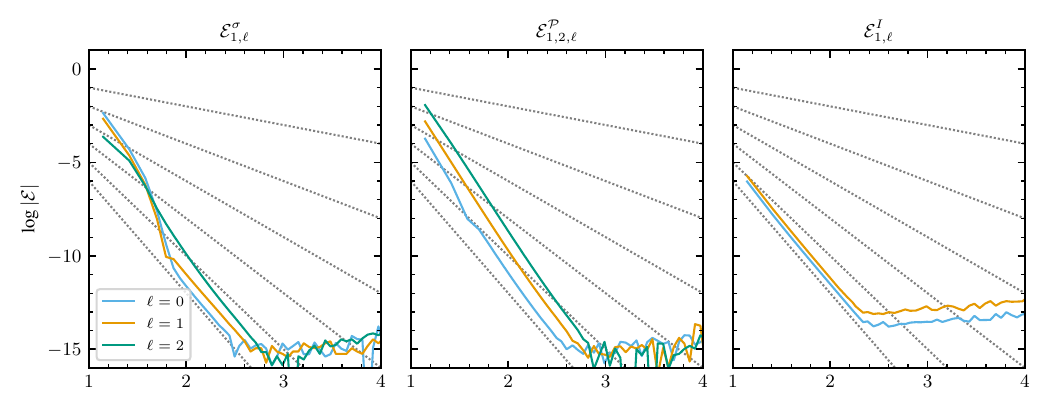}
  \caption{As in Fig.~\ref{f:anapoly-5-1-no-resolve-err}, except that
    a pair of sub-grids linked at the core-envelope boundary ($\xb
    \approx 0.577$) is used.} \label{f:anapoly-5-1-err}
\end{figure*}

The two polytrope models examined so far are characterized by a
smoothly varying interior structure. To explore the impact of sharp
structural features, we now consider a composite polytrope constructed
by matching analytic solutions for an $\npoly=5$ core and an
$\npoly=1$ envelope \citep[see][for details]{Eggleton:1998}. Although
density and pressure are continuous throughout this model, there is a
jump in the density gradient and hence the \bvtext\ frequency at the
core-envelope boundary, which we chose to place at radial coordinate
$\xb = 3^{-1/2} \approx 0.577$.

Figure~\ref{f:anapoly-5-1-no-resolve-err} reveals that the error
measures for this model behave quite differently than with the
previous ones. In each panel, the data are noisy but follow the trend
$\err \sim N^{-1}$. This slowed convergence is a consequence of the
\bvtext\ frequency jump. As shown by \citet{Gear:1984}, an order-$q$
discontinuity in the coefficients of an IVP introduces a local
$\order{h^q}$ error in numerical solutions that do not resolve the
discontinuity. In the present case $q=1$, leading to a $\sim N^{-1}$
contribution toward the error measures that quickly comes to dominate
other sources of error. The noise arises because certain specific
choices of $N$ place a grid point close to $\xb$, somewhat mitigating
the problem.

The fix here is simple: we split the grid into two sub-grids linked at
the core-envelope boundary; the first spans $x \in [0,\xb]$, the
second $x \in [\xb,1]$, and solutions are required to be continuous
across the boundary. Within each sub-grid, points are distributed as
described in Section~\ref{s:calcs}. Figure~\ref{f:anapoly-5-1-err}
demonstrates that this approach successfully eliminates the error from
the discontinuity, allowing the expected $\err \sim N^{-6}$ scalings
to emerge.

\subsection{Interpolated models} \label{s:interp-models}

\begin{figure*}[ht]
  \includegraphics{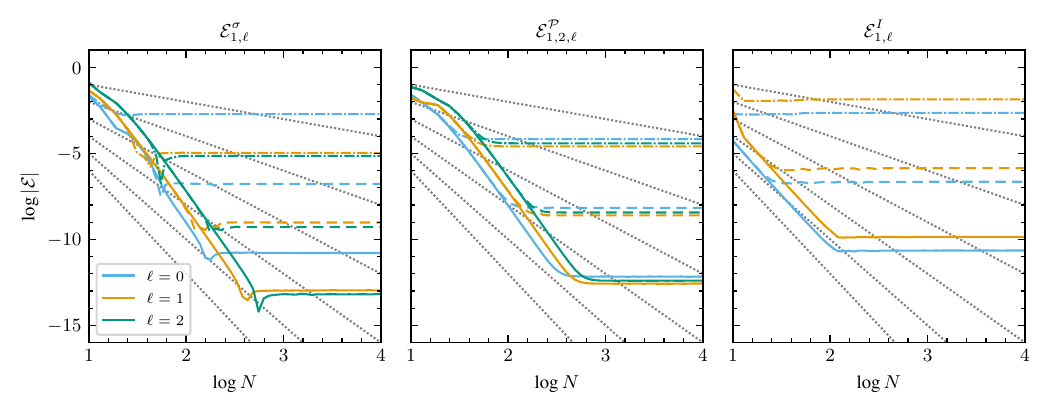}
  \caption{The $\errfreq[n,\ell]$, $\errorthog[n,n',\ell]$ and
    $\errfirst[n,\ell]$ error measures plotted as a function of the
    number of grid points $N$, for $(n,n')=(1,2)$, $\ell=0,1,2$ modes
    of the tabulated $\npoly=3$ polytrope. The number of tabulation
    points is $M=11$ (dot-dashed), $M=101$ (dashed) and $M=1001$
    (solid). The black dotted lines are the same as in previous
    figures.} \label{f:poly-3-err}
\end{figure*}

\begin{figure}[ht]
  \includegraphics{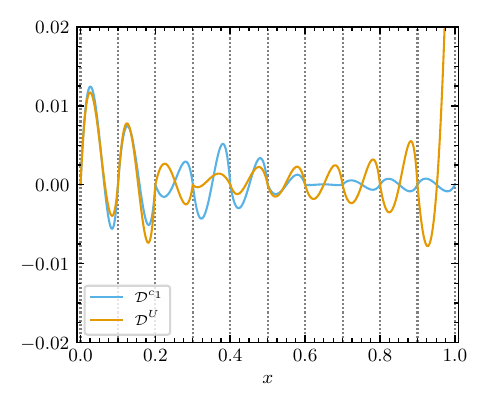}
  \caption{The defects $\defect[c_1]$ and $\defect[U]$ plotted as a
    function of fractional radius $x \equiv r/R$ for
    the tabulated $\npoly=3$ polytrope with $M=11$ points. The
    vertical dotted lines mark these points'
    abscissae.} \label{f:poly-3-interp-err}
\end{figure}

\begin{figure*}[ht]
  \includegraphics{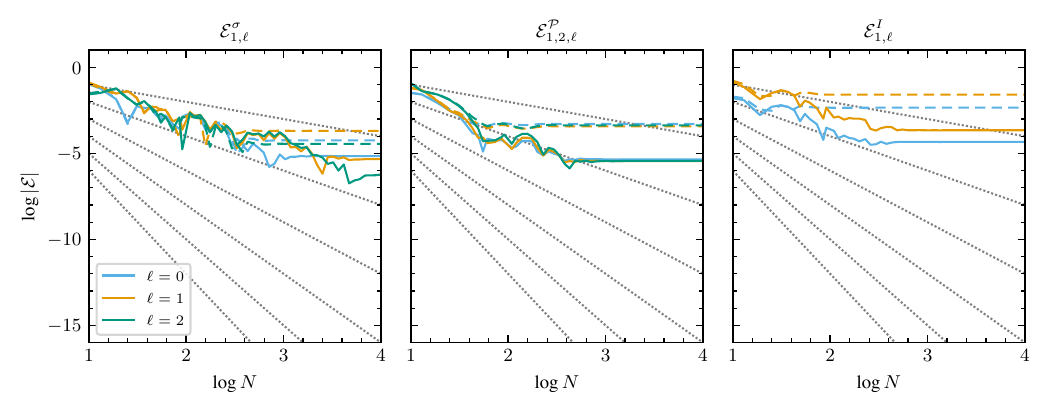}
  \caption{As in Fig.~\ref{f:poly-3-err}, but now showing results from
    the solar-like \mesa\ model. The number of tabulation points is
    $M=389$ (dashed) and $M=3621$ (solid).} \label{f:mesa-err}
\end{figure*}

\begin{figure}[ht]
  \includegraphics{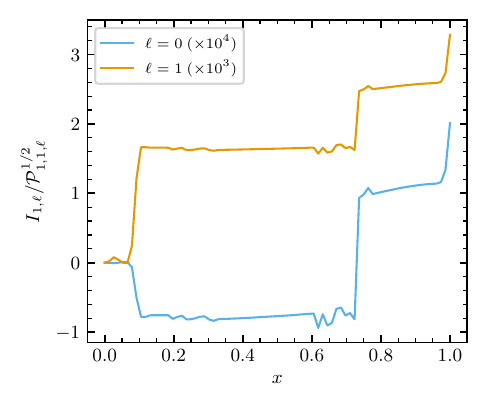}
  \caption{The first integral $I_{n,\ell}$ plotted as a function of
    fractional radius $x \equiv r/R$ for $n=1$, $\ell=0,1$ modes of the
    solar-like \mesa\ model with $M=3621$ tabulation
    points.} \label{f:mesa-first}
\end{figure}

\begin{figure}[ht]
  \includegraphics{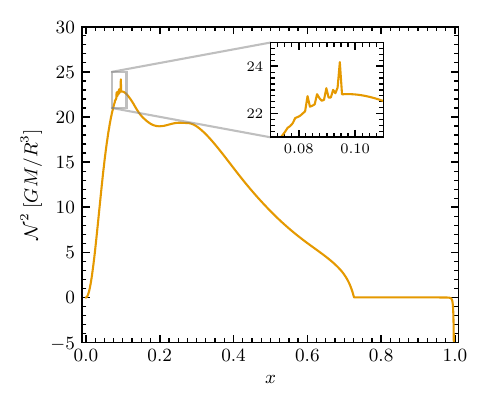}
  \caption{The square of the \bvtext\ frequency $\brunt^{2}$ plotted
    as a function of fractional radius $x \equiv r/R$ for the
    solar-like \mesa\ model with $M=3621$ tabulation
    points.} \label{f:mesa-brunt}
\end{figure}

The analytic stellar models considered so far are helpful for
illustrating the behavior of the error measures, but they are not
representative use-cases for oscillation codes. More typically, a
model is supplied to a code as a tabulation of various structural
quantities throughout the interior of the star. In general, the radial
grid of this tabulation does not match that used to discretize the
oscillation equations, and some kind of interpolation is therefore
required. These interpolated models are the focus of the present
section.

\subsubsection{$\npoly = 3$ polytrope} \label{s:poly-3-model}

We first consider an $\npoly=3$ polytrope tabulated on a uniform
$M$-point radial grid. The tabulation consists of three set of
values: $\{\polyz_{j}\}$ ($j = 1, 2, \ldots, M$) representing the
independent variable in the Lane-Emden equation, $\{\polyt_{j}\}$
representing the corresponding dependent variable, and
$\{\polyt'_{j}\}$ representing its first derivative. These data are
calculated by applying the \texttt{LSODAR} differential-equation
solver from \odepack\ \citep{Hindmarsh:1983} to the Lane-Emden
equation, with a relative tolerance parameter $10^{-15}$. They are
used by \gyre\ to construct piecewise cubic Hermite interpolants for
both $\polyt(\polyz)$ and $\polyt'(\polyz)$, allowing the stellar
structure to be evaluated at locations that fall between the
tabulation points.

Figure~\ref{f:poly-3-err} plots the three error measures as a function
of $N$, for tabulations composed of $M = 11$, 101 and 1001 points. All
three error measures initially decay rapidly, asymptoting toward the
expected $\err \sim N^{-6}$ scalings, before plateauing at constant
values. This stalled convergence is a consequence of the model
interpolation: although the cubic interpolants exactly satisfy the
Lane-Emden equation at the tabulation points, elsewhere they do
not. Therefore, the model is slightly inconsistent with the stellar
structure equations.

To illustrate the inconsistencies, we introduce the defects
\begin{equation} \label{e:defects}
  \begin{aligned}
  \defect[c_1] &\equiv \deriv{\ln c_{1}}{\ln x} - \left(3 - U\right), \\
  \defect[U] &\equiv \deriv{\ln U}{\ln x} - \left( - \frac{V}{\Gamma_{1}} - A^{\ast} - U + 3 \right)
  \end{aligned}
\end{equation}
(here, $V$, $A^{\ast}$, $U$ and $c_{1}$ are the coefficients appearing
in the dimensionless oscillation equations; see, e.g.,
\citealp{Unno:1989}). These defects are derived from equations~(20)
and (21), respectively, of \citet{Takata:2006a}, in such a way that
they both vanish when the equilibrium stellar structure locally
conserves mass. Figure~\ref{f:poly-3-interp-err} plots the defects as
a function of $x$, for the $\npoly=3$ polytrope tabulated with $M=11$
points. With the exception of the tabulation points (where
$\defect[c_1] = \defect[U] = 0$ by construction), it is generally the
case that the defects are non-zero, highlighting the breakdown of mass
conservation.

The net effect of this breakdown is to introduce additional terms on
the right-hand side of the error
measures~(\ref{e:error-freq}--\ref{e:error-first}), that are
independent of $N$ but decrease with increasing $M$ because finer
tabulations reduce the magnitude of both $\defect[c_1]$ and
$\defect[U]$. For the data shown in the figure, switching from $M=11$
to $M=101$, and likewise from $M=101$ to $M=1001$, yields a $\approx
10^{-4}$ reduction in the error-measure plateaus, indicating that the
errors from interpolation inconsistency scale as $\sim M^{-4}$.

\subsubsection{\mesa\ model} \label{s:mesa-model}

We now consider models constructed using release 24.03.01 of the
\mesa\ stellar evolution code
\citep{Paxton:2011,Paxton:2013,Paxton:2015,Paxton:2018,Paxton:2019,Jermyn:2023}. We
evolve a \SI{1}{\Msun} solar-metallicity star from the zero-age main
sequence until its luminosity reaches \SI{1}{\Lsun}, providing a
simple approximation to the present-day Sun. We repeat this exercise
for two different choices of \mesa's grid refinement parameter,
\texttt{mesh\_delta\_coeff} = 2.0 and \texttt{mesh\_delta\_coeff} =
0.2, leading to tabulation with $M = 389$ and $M = 3621$ points,
respectively. From these data, \gyre\ constructs piecewise cubic
monotonic interpolants \citep{Steffen:1990} for $V_{2} \equiv x^{-2}
V$, $A^{\ast}$, $U$, $c_{1}$ and $\Gamma_{1}$.

Figure~\ref{f:mesa-err} repeats the $\npoly=3$ calculations for the
two \mesa\ models. The error measures exhibit the same stalled
convergence as in Fig.~\ref{f:poly-3-err}, which again is a
consequence of the model interpolation (although now the error-measure
plateau scales as $\sim M^{-2}$). However, prior to the stall the
error measures decrease as $\sim N^{-2}$, albeit with significant
noise. This behavior is reminiscent of
Fig.~\ref{f:anapoly-5-1-no-resolve-err}, where the unresolved
structural discontinuity in the composite polytrope results in slowed
convergence. To examine what features in the \mesa\ models might be
responsible for this behavior, Fig.~\ref{f:mesa-first} plots the first
integrals $I_{n,\ell}$ as a function of $x$ for the $n=1$, $\ell=0,1$
modes of the $M=389$ model, discretized on a grid with $N=77$
points. At three locations $x \approx 0.09, 0.73, 0.99$ the first
integrals change rapidly, resulting in significant contributions
toward the $\errfirst[n,\ell]$ error measure. Examining the model
structure, each of these locations can be identified with features in
the \bvtext\ frequency profile (plotted in Fig.~\ref{f:mesa-brunt})
that prove challenging to \gyre:
\begin{itemize}
\item At $x = 0.09$ there are small-scale oscillations in
  $\brunt^2$. These arise from corresponding fluctuations in the
  gradient of the mean molecular weight, produced by the retreating
  boundary of a transient convective core during early main-sequence
  evolution.
\item At $x = 0.73$, there is an abrupt change in the gradient of
  $\brunt^2$. This arises from transition between the radiative
  interior ($\brunt^2 > 0$) and convective envelope ($\brunt^{2}
  \approx 0$).
\item At $x = 0.99$, $\brunt^2$ grows rapidly to large, negative
  values. This arises from the transition to inefficient, highly
  super-adiabatic convection in the surface layers.
\end{itemize}
Each of these features appear to \gyre\ as order-2 discontinuities in
$A^{\ast} \equiv r g^{-1} \brunt^{2}$, and therefore introduces
additional contributions $\sim N^{-2}$ toward the error measures that
result in the slowed convergence seen in Fig.~\ref{f:mesa-err}. In
principle, a judicious choice of grid-point distribution can eliminate
these contributions, as was successfully done with the composite
polytrope in Section~\ref{s:poly-5-1-model}. However, while an
optimized grid may accelerate the convergence of the error measures at
small $N$, it has no effect on the plateaus seen in
Fig.~\ref{f:mesa-err} toward larger $N$. We confirm this by repeating
the calculations using \gyre's built-in grid-refinement strategies,
which place grid points based on the behavior of both the equilibrium
stellar structure and the eigenfunctions. This results in no
significant reduction in the error measure plateaus. We are therefore
left with the conclusion that \emph{only} way to push down these
plateaus is to use models with finer (larger-$M$) tabulations.


\section{Summary \& Discussion} \label{s:discuss}

In the previous sections we introduce three measures,
$\errfreq[n,\ell]$, $\errorthog[n,n',\ell]$, and $\errfirst[n,\ell]$,
as tools for characterizing the numerical error in solutions to the
oscillation equations. For a variety of stellar models, we explore how
these error measures respond to changes in the number of radial grid
points $N$ used to discretize the oscillation equations, and/or the
number of points $M$ in the tabulation of the stellar structure. To
summarize the results from this investigation, we propose an idealized
model for the behavior of the error measures as a function of $N$ and
$M$:
\begin{equation} \label{e:error-model}
  |\err| \approx \adisc N^{-\bdisc} + \aquad N^{-\bquad} + \acont N^{-\bcont} + \arandom N + \ainterp M^{-\binterp}.
\end{equation}
Each term on the right-hand side represents a particular source of
numerical error, with the $a$ coefficients setting the magnitude of
the error and the $b$ coefficients determining its scaling with $N$ or
$M$. In detail:
\begin{itemize}
  \item the $\adisc N^{-\bdisc}$ term represents the truncation error
    arising from the discretization of the oscillation equations. For
    a discretization scheme with local error $\order{h^{i+1}}$, $\bdisc = i$.
  \item the $\aquad N^{-\bquad}$ term represents the truncation error
    arising from the quadrature used to evaluate the integral terms in
    the definitions of $\errfreq[n,\ell]$ and
    $\errorthog[n,n',\ell]$. For a quadrature scheme with local error
    $\order{h^{i+1}}$, $\bquad = i$. The $\errfirst[n,\ell]$ error
    measure does not require quadrature, and so has $\aquad = 0$.
  \item the $\acont N^{-\bcont}$ term represents the error arising
    from the lowest-order discontinuity (or discontinuities) in the
    equilibrium stellar structure. If this discontinuity has an order
    $q$, then $\bcont = q$.
  \item the $\arandom N$ term represents the accumulated round-off
    error.
  \item the $\ainterp M^{-\binterp}$ term represents the
    inconsistencies arising from interpolation of tabulated
    models. Evaluated models do not require interpolation, and so have
    $\ainterp = 0$.
\end{itemize}
This model does not account for calculation \emph{mistakes} --- for
instance, neglecting non-zero surface terms when evaluating
$\errfreq[n,\ell]$ (see the dot-dashed curves in the left-hand panels
of Figs.~\ref{f:anapoly-5-err}). Moreover, it is intended to be
descriptive rather than prescriptive --- a way of understanding the
behavior of error measures, rather than predicting this behavior
\emph{a priori}.

Although our calculations focus exclusively on the \gyre\ code, other
oscillation codes can benefit from a similar exploration of the error
measures' functional dependence on $N$ and $M$. In addition to
verifying basic code correctness, such exercises can furnish a
mechanism for selecting $N$, $M$ values that strike an appropriate
balance between computational cost and accuracy. In this context, we
expect smaller $\err$ to correspond to more-accurate solutions;
however, to place this connection on a more quantitative footing, it
behooves us to consider the semantics of each error measure --- what
actually is being measured?

To clarify the meaning of the frequency error measure, we write the
$\sigmae_{n,\ell}$ and $\sigmai_{n,\ell}$ terms appearing on the
right-hand side of equation~(\ref{e:error-freq}) as the sum of the
exact frequency $\sigma_{n,\ell}$ and a numerical error:
\begin{equation}
  \sigmae_{n,\ell} = \sigma_{n,\ell} + \Delta \sigmae_{n,\ell}, \qquad
  \sigmai_{n,\ell} = \sigma_{n,\ell} + \Delta \sigmai_{n,\ell}.
\end{equation}
Then, equation~(\ref{e:error-freq}) becomes
\begin{equation} \label{e:errfreq-delta}
  \errfreq[n,\ell] = \frac{\Delta \sigmae_{n,\ell} - \Delta \sigmai_{n,\ell}}{\sigma_{n,\ell}}
\end{equation}
where we have neglected terms that have a combined error order of
second or beyond. From this, we can establish the inequality
\begin{equation} \label{e:errfreq-inequal}
  | \errfreq[n,\ell] | \leq 2 \max \left(
  \left| \frac{\Delta \sigmae_{n,\ell}}{\sigma_{n,\ell}} \right|,
  \left| \frac{\Delta \sigmai_{n,\ell}}{\sigma_{n,\ell}} \right|
  \right),
\end{equation}
and so the frequency error measure sets a lower bound on the magnitude
of the relative error in $\sigmae_{n,\ell}$ or $\sigmai_{n,\ell}$,
whichever is greater. Turning this statement around, if it is required
that the relative error in the eigenfrequency be less than some
threshold $E$, then it is necessary that $|\errfreq[n,\ell]| \leq
2E$. Unfortunately, this condition is not also sufficient;
circumstances situations can potentially arise where
$|\errfreq[n,\ell]|$ is significantly smaller than $E$, but the
relative errors in both $\sigmae_{n,\ell}$ and $\sigmai_{n,\ell}$
remain larger than $E$.

The situation is somewhat improved in cases where the surface term
satisfies $\surf[n,n',\ell] = \surf[n',n,\ell]^{\ast}$ --- for
instance, due to adoption of the zero-pressure boundary
condition~(\ref{e:zero-press-bound}). The variational principle then
predicts that $(\Delta \sigmai_{n,\ell} / \sigma_{n,\ell}) \sim
(\Delta \sigmae_{n,\ell} / \sigma_{n,\ell})^{2}$. In the limit
$|\Delta \sigmae_{n,\ell} / \sigma_{n,\ell} | \ll 1$, the second term
in the numerator of equation~(\ref{e:errfreq-delta}) can be neglected,
and equation~(\ref{e:errfreq-inequal}) then reduces to
\begin{equation}
  | \errfreq[n,\ell] | \approx \left| \frac{\Delta \sigmae_{n,\ell}}{\sigma_{n,\ell}} \right|,
\end{equation}
indicating that the frequency error measure \emph{is} the relative
error in the eigenfrequency. However, a word of caution here
\citep[echoing the remarks in section 4
  of][]{Christensen-Dalsgaard:1979}: $\Delta \sigmai_{n,\ell}$
includes a contribution from the truncation error that arises when
evaluating the integrals in equation~(\ref{e:freq}), and this
contribution is not subject to the variational principle. Therefore,
neglecting $\Delta \sigmai_{n,\ell}$ requires that a sufficiently
accurate quadrature scheme be used.

Interpreting the orthogonality error measure is rather more
straightforward: it represents the cosine of the angle in the Hilbert
space defined by the inner products $\proj[n,n',\ell]$. This
information is useful when expanding solutions to certain problems in
a superposition of eigenfunctions --- for instance, when calculating
the response of a star to tidal forcing by a companion object
\citep[see, e.g.,][]{Burkart:2012,Fuller:2012}. Then,
$\errorthog[n,n',\ell]$ can be used to quantify the level of numerical
leakage between terms in the expansion.

The meaning of the first-integral error measure is perhaps the most
elusive. The first integrals derived by \citet{Takata:2006a} should
vanish everywhere when a star is isolated from any outside
gravitational influence. Therefore, $\errfirst[n,\ell] \neq 0$
indicates the contamination of numerical eigenfunctions with
additional perturbations that arise from an external distribution of
mass. Knowing what this distribution is seems to have little practical
value; however, knowing \emph{where} in the star the resulting
contamination occurs is certainly useful. We demonstrate this in
Figs.~\ref{f:mesa-first} and Fig.~\ref{f:mesa-brunt}, where the jumps in the
$I_{0}$ and $I_{1}$ first integrals allows us swiftly to diagnose
features in the $\brunt^{2}$ profile that are causing difficulties for
\gyre.

With these considerations, we return now the question of selecting
appropriate values for $N$ and $M$. For asteroseismic optimization
studies \citep[e.g.,][]{Aerts:2018}, where the goal is to construct a
(tabulated) stellar evolutionary model whose numerical
eigenfrequencies match a set of observed frequencies to within some
relative tolerance $E$, we propose the following course of action:
\begin{enumerate}[(i)]
\item For a model with a given $M$, create plots of $\errfreq[n,\ell]$
  as a function of $N$ for modes of interest, and identify the regimes
  where the various terms in equation~(\ref{e:error-model}) are
  dominant.
\item If interpolation errors mean that $|\errfreq[n,\ell]| > E$ for
  all $N$ considered, then construct a new model with a larger $M$ and
  repeat step (i). As part of this process, inspecting the defects
  $\defect[c_1]$, $\defect[U]$ defined in equation~(\ref{e:defects})
  may prove useful in identifying where additional tabulation points
  should be placed.
\item If $|\errfreq[n,\ell]| < E$ once $N$ exceeds some threshold
  $N_{\rm thr}$, then adopt the eigenfrequencies at $N_{\rm th}$. If a
  smaller $N_{\rm th}$ is desired (for instance, in the interests of
  computational efficiency), then step (i) can be repeated using
  higher-order discretization or quadrature schemes, or a distribution
  of grid points that's optimized to resolve discontinuities (the
  choice here depending on which error term is dominant). Richardson
  extrapolation \citep[e.g.,][]{Press:1992,Christensen-Dalsgaard:1994}
  can also be applied in tandem with these approaches, to achieve
  further reductions in $N_{\rm th}$.
\end{enumerate}
Step (ii) will likely be the costliest in this procedure, as the
computational cost of stellar evolution calculations is usually much
larger than oscillation calculations. Therefore, future work should
prioritize finding ways (other than increasing $M$) to reduce the
impact of interpolation errors --- for instance, by devising a set of
interpolants that are locally consistent with the stellar structure
equations.

There also remains a need to examine the influence of outer boundary
conditions on the error measures. Our choice of the zero-pressure
mechanical boundary condition~(\ref{e:zero-press-bound}) and the
potential condition~(\ref{e:pot-bound}) is primarily motivated by
simplicity, but also leads to favorable mathematical outcomes (e.g.,
that $\surf[n,n',\ell] = \surf[n',n,\ell]^{\ast}$, implying the
orthogonality of eigenfunctions). However, this choice may sacrifice
some degree of physical realism: stars possess atmospheres that extend
beyond their formal surfaces, and boundary conditions should
adequately account for the fact that oscillations can and do penetrate
out into these superficial layers.

In this regard, \citet[][their section 18.1]{Unno:1989} present a
more-general formulation of the surface boundary conditions, that
approximates the oscillation equations in the atmosphere as having
spatially constant coefficients\footnote{The older
\citet{Dziembowski:1971} boundary condition results from expanding
this formulation to first order in $V^{-1}$.}. When adopting this
formulation, it is appropriate to extend the upper bounds of the
$\weight[n,n',\ell]$ and $\proj[n,n',\ell]$ integrals
(equations~\ref{e:weight} and~\ref{e:proj}) to include the atmosphere;
this can be done by leveraging the explicit solutions to the
constant-coefficient equations \citep[see, e.g., equation~18.36
  of][]{Unno:1989}. However, we anticipate that additional
contributions to the error measures will arise due to the
approximations made in treating the atmosphere.  We plan to explore
this matter in detail in a future paper.

On a closing note, we highlight that the tools and analysis presented
here address only one part of the asteroseismic code verification
problem. It is equally important to explore the numerical errors
arising in the stellar evolution calculations that feed into
oscillation codes. The recent work by \citet{Li:2025} is an important
step in this direction.


\section*{Acknowledgments}

We thank the anonymous referee for their insightful remarks during the
reviewing process. RHDT is immensely grateful for the hospitality of
the Center for Computational Astrophysics during his time there as a
sabbatical visiting researcher. RHDT and RVK both acknowledge support
from NASA grants 80NSSC20K0515, 80NSSC23K1517 and 80NSSC24K0895.

\facilities{We have made extensive use of NASA's
  Astrophysics Data System Bibliographic Services.}

\software{%
  Astropy \citep{astropy:2013,astropy:2018,astropy:2022},
  \gyre\ \citep{Townsend:2013,Townsend:2018,Goldstein:2020,Sun:2023},
  Matplotlib \citep{Hunter:2007},
  \mesa\ \citep{Paxton:2011,Paxton:2013,Paxton:2015,Paxton:2018,Paxton:2019,Jermyn:2023}.
  Sample input files for \gyre\ and \mesa\ are available on Zenodo
  under an open-source Creative Commons Attribution license:
  \dataset[10.5281/zenodo.15399940]{https://doi.org/10.5281/zenodo.15399940}.
}


\bibliography{paper}


\newpage

\appendix

\section{Mathematical Formalism} \label{a:formal}

In this appendix we lay out the formalism that provides the basis for
the error measures introduced in Section~\ref{s:error}. The goal is to
provide sufficient detail that there can be no ambiguity in their
definition. Throughout, we assume a context of linear, adiabatic
oscillations of spherical, non-rotating stars.

\subsection{Perturbation Forms} \label{a:pert-forms}

The linearized fluid equations describing small departures from the
equilibrium state of a star can be separated in all three
spherical-polar coordinates $(r,\theta,\phi)$ and in time $t$. For a
mode with radial order $n$, harmonic degree $\ell$ and azimuthal order
$m$, we express the fluid displacement perturbation vector in the form
\begin{equation} \label{e:pert-xi}
  \vxi_{n,\ell,m}(\vpos;t) = \left\{
      \txir[n,\ell](r) \, \ver +
      \txih[n,\ell](r) \left[
        \vetheta \, \pderiv{}{\theta} +
        \vephi \, \frac{1}{\sin\theta} \, \pderiv{}{\phi}
        \right]
      \right\} \,
    Y^{m}_{\ell}(\theta,\phi) \,
    \exp{(-\ii\,  \sigma_{n,\ell} \, t)}.
\end{equation}
Here, $\vpos$ denotes the position vector; $(\ver,\vetheta,\vephi)$
are the spherical-polar basis vectors at $\vpos$; and $Y^{m}_{\ell}$
is a spherical harmonic. The accompanying Eulerian perturbations to
the pressure $P$ and self-gravitational potential $\Phi$ take the form
\begin{equation} \label{e:pert-other}
  \begin{aligned}
  P'_{n,\ell,m}(\vpos;t) &= \tP[n,\ell]'(r) \, Y^{m}_{\ell}(\theta,\phi) \, \exp{(-\ii\, \sigma_{n,\ell} \, t)}, \\
  \Phi'_{n,\ell,m}(\vpos;t) &= \tPhi[n,\ell]'(r) \, Y^{m}_{\ell}(\theta,\phi) \, \exp{(-\ii\, \sigma_{n,\ell} \, t)},
  \end{aligned}
\end{equation}
respectively.

In these expressions, the functions with tilde accents
($\txir[\ell,n], \tP[\ell,n]'$, etc.) encapsulate the radial
dependencies of perturbations. They are the eigenfunctions of a
two-point boundary value problem (BVP) constructed from a coupled
system of ordinary differential equations and boundary conditions
\citep[see, e.g., section 14.1 of][]{Unno:1989}; the corresponding
eigenvalue is $\sigma_{n,\ell}^{2}$, with $\sigma_{n,\ell}$ being the
angular frequency of the oscillation mode.

\subsection{Integral Expression for Mode Frequencies} \label{a:freq}

As demonstrated by many authors \citep[see,
  e.g.,][]{Epstein:1950,Chandrasekhar:1964,Goossens:1974,Schwank:1976,Cox:1980,Christensen-Dalsgaard:1982,Kawaler:1985,Unno:1989},
the frequency of an oscillation mode can be expressed in terms of
integrals over the mode's eigenfunctions. Most derivations of this
relationship make specific assumptions about the boundary conditions
applied at the stellar surface; here, we present a somewhat
more-general approach.

We begin from equation~(14.18) of \citet{Unno:1989}, which originates
in the linearized momentum equation. Adapting this equation to our
notation, and performing the integrals over $\theta$ and $\phi$, we
obtain the relationship
\begin{equation} \label{e:integrated-mom}
  \sigma_{n,\ell}^{2} \, \proj[n,n',\ell] = \weight[n,n',\ell] + \surf[n,n',\ell],
\end{equation}
where
\begin{equation} \label{e:proj}
  \proj[n,n',\ell] =
  \int_{0}^{R} \left[
    \txir[n,\ell] \, \txir[n',\ell]^{\ast} +
    \ell(\ell+1) \, \txih[n,\ell] \, \txih[n',\ell]^{\ast}
    \right] \, \rho \, r^{2} \, \diff{r},
\end{equation}
defines an inner product between the displacement perturbation eigenvectors
for a pair of modes with indices $(n,\ell)$ and $(n',\ell)$,
\begin{equation} \label{e:weight}
  \weight[n,n',\ell] =   
  \int_{0}^{R} \left\{
  \frac{1}{\rho^{2} c^{2}} \tP[n,\ell]' \tP[n',\ell]'^{\ast} + 
  \brunt^{2} \, \txir[n,\ell] \txir[n',\ell]^{\ast} -
  \frac{1}{4\pi G \rho}
  \left[ \pderiv{\tPhi[n,\ell]'}{r} + \frac{\ell+1}{r} \tPhi[n,\ell]' \right]
  \left[ \pderiv{\tPhi[n',\ell]'}{r} + \frac{\ell+1}{r} \tPhi[n',\ell]' \right]^{\ast}
  \right\} \, \rho \, r^{2} \, \diff{r},
\end{equation}
defines a generalized weight integral\footnote{We used this
terminology because, when $n'=n$, the integral reduces to the standard
weight integrals appearing in the literature
(\citealp[e.g.,][]{Kawaler:1985}, as corrected by
\citealp{Townsend:2023})} for the same pair of modes, and
\begin{equation} \label{e:surf-term}
  \surf[n,n',\ell] = \left\{ r^{2} \left[
    \left( \tP[n,\ell]' + \rho \, \tPhi[n,\ell]' \right) \txir[n',\ell]^{\ast} +
    \frac{1}{4\pi G} \tPhi[n,\ell]' \left( \deriv{\tPhi[n',\ell]'}{r} + \frac{\ell+1}{r} \tPhi[n',\ell]' \right)^{\ast}
    \right] \right\}_{r=R},
\end{equation}
is a surface term, with $R$ the stellar radius. In these expressions,
\begin{equation} \label{e:sound-brunt}
  c^{2} \equiv \frac{\Gamma_{1} P}{\rho}, \qquad
  \brunt^{2} \equiv \frac{g}{r} \left( \frac{1}{\Gamma_{1}} \deriv{\ln P}{\ln r} - \deriv{\ln \rho}{\ln r} \right),
\end{equation}
are the squares of the adiabatic sound speed and \bvtext\ frequency,
respectively, $\rho$ is the density, $g$ the gravitational
acceleration, and $\Gamma_{1}$ the first adiabatic exponent.

Setting $n'=n$ in equation~(\ref{e:integrated-mom}) yields, after a
little rearrangement,
\begin{equation} \label{e:freq}
  \sigma_{n,\ell} = \left[\frac{\weight[n,n,\ell] + \surf[n,n,\ell]}{\proj[n,n,\ell]}\right]^{1/2}
\end{equation}
which constitutes our formulation of the integral expression for mode
frequencies. Other formulations are possible, which in certain
circumstances may exhibit better numerical behavior \citep[see, for
  instance, the discussion in appendix D
  of][]{Christensen-Dalsgaard:1982}.

The integrals appearing in equations~(\ref{e:proj})
and~(\ref{e:weight}) adopt the stellar radius $R$ as the upper
limit. However, we note that the integral expression~(\ref{e:freq})
remains valid for any other choice of upper limit, as long as the
surface term $\surf[n,n,\ell]$ is evaluated at the same location.

\subsection{Orthogonality of Eigenfunctions} \label{a:orthog}

We now consider the conditions under which oscillation eigenfunctions
are mutually orthogonal. Observing that $\proj[n',n,\ell] =
\proj[n,n',\ell]^{\ast}$ and $\weight[n',n,\ell] =
\weight[n,n',\ell]^{\ast}$, it follows from
equation~(\ref{e:integrated-mom}) that
\begin{equation} \label{e:freq2-diff}
  \left( \sigma_{n,\ell}^{2} - \sigma_{n',\ell}^{2\ast} \right) \proj[n,n',\ell] = \surf[n,n',\ell] - \surf[n',n,\ell]^{\ast}.
\end{equation}
When $n'=n$, this reduces to
\begin{equation}
  \left( \sigma_{n,\ell}^{2} - \sigma_{n,\ell}^{2\ast} \right) \proj[n,n,\ell] = \surf[n,n,\ell] - \surf[n,n,\ell]^{\ast}.
\end{equation}
If $\imag(\surf[n,n,\ell]) = 0$, then the right-hand side of this
equation vanishes,
\begin{equation}
  \left( \sigma_{n,\ell}^{2} - \sigma_{n,\ell}^{2\ast} \right) \proj[n,n,\ell] = 0;
\end{equation}
given that non-trivial solutions have $\proj[n,n,\ell] > 0$, this
indicates that the eigenvalues $\sigma_{n,\ell}^{2}$ are real.

Returning to equation~(\ref{e:freq2-diff}), if $\surf[n,n',\ell] =
\surf[n',n,\ell]^{\ast}$ then
\begin{equation} 
  \left( \sigma_{n,\ell}^{2} - \sigma_{n',\ell}^{2} \right) \proj[n,n',\ell] = 0.
\end{equation}
Assuming that no degeneracies exist among the eigenvalues \citep[see
  section 14.1 of][for a brief discussion of this
  assumption]{Unno:1989}, the term in parentheses is non-zero when $n'
\neq n$ and so
\begin{equation} \label{e:orthog}
  \proj[n,n',\ell] = 0,
\end{equation}
indicating that eigenfunctions of modes with differing radial orders
but the same harmonic degree are orthogonal to each other. This is
equivalent to the statement that the oscillation BVP is self-adjoint,
and leads to the well-known result
\citep[e.g.,][]{Chandrasekhar:1964,Lynden-Bell:1967} that a
variational approach can be used to refine its eigenvalues.

With the potential boundary condition~(\ref{e:pot-bound}), the
condition $\surf[n,n',\ell] = \surf[n',n,\ell]^{\ast}$ necessary for
orthogonality simplifies to
\begin{equation} \label{e:orthog-with-pot-bound}
  \left( \tP[n,\ell]' \, \txir[n',\ell]^{\ast} \right)_{r=R} =  
  \left( \tP[n',\ell]^{\prime\ast} \, \txir[n,\ell] \right)_{r=R}
\end{equation}
This is satisfied by the zero-pressure boundary
condition~(\ref{e:zero-press-bound}), but can also be satisfied by
other boundary conditions such as the rigid-lid condition $\txir[n,\ell]=0$.

\end{document}